\documentclass[aps,prfluids,preprint,unsortedaddress]{revtex4-2}
\usepackage{graphicx}
\usepackage{epstopdf, epsfig}
\usepackage{amsmath}
\usepackage{amsfonts}

\usepackage{xcolor}
\usepackage{caption}
\usepackage{subcaption}
\usepackage{tikz}
\usetikzlibrary{shapes}
\usepackage[absolute,overlay]{textpos}
\usepackage[colorlinks=true,allcolors=blue]{hyperref}
\usetikzlibrary{shapes.geometric}

\newcommand{\markertri}{\raisebox{0.5pt}{\tikz{\node[draw,scale=0.4,regular polygon,regular polygon sides=3, rotate=0,fill=white](){};}}}
\newcommand{\markertridown}{\raisebox{0.5pt}{\tikz{\node[draw,scale=0.4,regular polygon,regular polygon sides=3, rotate=180,fill=white](){};}}}
\newcommand{\markersquare}{\raisebox{0.2pt}{\tikz{\node[draw,scale=0.6,regular polygon,regular polygon sides=4, rotate=0,fill=white](){};}}}
\newcommand{\markercircle}{\raisebox{0.2pt}{\tikz{\node[draw,scale=0.6,circle,fill=white](){};}}}
\newcommand{\markerdiamond}{\raisebox{0.5pt}{\tikz{\node[draw,scale=0.5,diamond,fill=white](){};}}}
\newcommand{\markerstar}{\raisebox{0.5pt}{\tikz{\node[draw,scale=0.5,star,fill=white](){};}}}

\newcommand*\vb[1]{\boldsymbol{#1}}
\newcommand*\mb[1]{\mathbf{#1}}
\newcommand*\norm[1]{\lVert #1 \rVert}
\newcommand{\me}{\mathrm{e}}
\newcommand{\St}{\mathit{St}}
\newcommand*\change[1]{\textcolor{black}{#1}}
\usepackage{algpseudocode}
\newcounter{algorithm}
\renewcommand{\thealgorithm}{\arabic{algorithm}}

\newenvironment{algorithmfloat}[1][tb]
{%
  \refstepcounter{algorithm}% increment counter
  \par\medskip
  \noindent
  \centering
  \begin{minipage}{\textwidth} % 80% of text width
  {ALGORITHM \thealgorithm.} #1
  \par\medskip
  \hrule\vspace{1mm}\hrule\vspace{1mm}
  \noindent
}
{%
\vspace{1mm}\hrule\vspace{1mm}\hrule\vspace{1mm}
\end{minipage}
  \par\medskip
}
\begin{document}

\title{Spectral proper orthogonal decomposition of rapid snapshot pairs sampled at sub-Nyquist intervals}
% repeat the \author .. \affiliation  etc. as needed

\author{Caroline Cardinale}
\email{Contact author: ccardina@caltech.edu}

\affiliation{Department of Civil and Mechanical Engineering, California Institute of Technology, Pasadena, CA, 91125, USA}
\author{Steven L. Brunton}
\affiliation{Department of Mechanical Engineering, University of Washington, Seattle, WA, 98195, USA}
\author{Tim Colonius}
%\email[]{Your e-mail address}
%\homepage[]{Your web page}
%\thanks{}
%\altaffiliation{}
\affiliation{Department of Civil and Mechanical Engineering, California Institute of Technology, Pasadena, CA, 91125, USA}

\date{\today}

\begin{abstract}
% insert abstract here
Modal decomposition methods are important for characterizing the low-dimensional dynamics of complex systems, including turbulent flows. Different methods have varying data requirements and produce modes with different properties. 
Spectral proper orthogonal decomposition (SPOD) produces orthogonal, energy-ranked spatial modes at discrete temporal frequencies for statistically stationary flows.  However, SPOD requires long stretches of sequential, uniformly sampled, time-resolved data. These data requirements limit SPOD's use in experimental settings where the maximum capture rate of a camera is often slower than the Nyquist sampling rate required to resolve the highest turbulent frequencies. However, if two PIV systems operate in tandem, pairs of data can be acquired that are arbitrarily close in time. The dynamic mode decomposition (DMD) uses this {\it pairwise data} to resolve frequencies up to the Nyquist frequency associated with the small time step within a pair. However, these modes do not form a basis and have no set ranking. The present work attempts to compute SPOD modes from pairwise data with a small time step but with large gaps between pairs.  We use DMD on pairwise data to estimate segment-wise, uniformly sampled series that can then be used to estimate the SPOD modes, intending to resolve frequencies between the gap and pair Nyquist limits. The method is tested on numerically obtained data of the linearized complex Ginzburg-Landau equation, as well as a Mach 0.4 isothermal turbulent jet. For the jet, pairwise SPOD can accurately de-alias the SPOD spectrum and estimate mode shapes at frequencies up to $St \approx 1.0$ while using over 90\% less data.
\end{abstract}

% insert suggested keywords - APS authors don't need to do this
%\keywords{}

%\maketitle must follow title, authors, abstract, and keywords
\maketitle

% body of paper here - Use proper section commands
% References should be done using the \cite, \ref, and \label commands
\section{Introduction}
Proper orthogonal decomposition (POD)~\citep{lumley1967structure,lumley1970stochastic} provides an optimal, energy-ranked basis for describing spatio-temporally correlated structures in a data set.  Space-only formulations are often applied, although they disregard temporal information. In contrast, for statistically stationary data, spectral POD (SPOD) yields modes at discrete frequencies that best represent the second-order space-time statistics~\citep{towne2018spectral}. However, SPOD requires a long stretch of uniformly sampled, time-resolved data; here time-resolved means that the data is sampled faster than the Nyquist sampling rate required to resolve the highest frequencies present in the flow. In a computational setting, this uniformly sampled, time-resolved data is available, although memory intensive. In experiments, however, time-resolved particle image velocimetry (PIV) remains challenging, particularly in high-speed flows. If a camera is too slow, aliasing contaminates the SPOD spectrum.

Dynamic mode decomposition (DMD)~\citep{Rowley2009jfm,schmid2010dynamic,schmid2022dynamic}, produces temporally coherent spatial modes that oscillate, grow, and decay at specific rates in time. These modes are not optimal \change{and do not form a basis,} but variants of the algorithm, such as exact DMD~\citep{tu2014dmd}, are available to address non-uniformly sampled data.  \change{Exact DMD can be applied to two sequences of data that are offset by a constant timestep; each sequence need not be uniformly sampled or sampled at a rate higher than the Nyquist frequency. Combined as a single time series, these two sequences are pairs of data separated by long gaps. \citet{bridges2003measurements} proposed a method to obtain the described pairwise data using two PIV setups specifically for looking at the space-time correlations in turbulent jets; the time offset between the two sequences can then be made arbitrarily small.}  Despite long gaps between pairs, DMD can resolve frequencies up to the Nyquist frequency associated with the time step between the two images of the pair, rather than the restricted sampling rate limited by the speed of a single PIV setup~\citep{tu2014dmd}.  This allows one to observe the highest frequencies in turbulent flows despite camera speeds being limited to the order of kHz; most PIV setups are restricted to even lower frequencies~\citep{taira2017modal}.  Additionally, in PIV, there is a trade-off between camera capture rate and camera resolution. Thus the pairwise approach allows for higher spatial resolution by decreasing the sampling rate of each camera.

In this paper, we propose an algorithm that leverages exact DMD to estimate SPOD for pairwise data.  There have been related previous attempts to perform spectral analysis when time-resolved data is not available.~\citet{tu2014spectral} leveraged sparsity in the frequency domain to estimate the power spectrum and DFT modes of flow past a cylinder where the time between snapshots was randomly varied between a user-set range. This work showed the efficacy of random sampling as a way to beat the typical Nyquist criterion.~\citet{nekkanti2023gappy} developed gappy SPOD to estimate corrupted or missing regions of snapshots. At $20\%$ data loss, their method recovered 97\% of the missing regions. However, even one missing snapshot between pairs ($\tau=3\Delta t$) results in a 33\% loss of data, a case for which gappy SPOD has not been tested. There have also been several attempts to combine non-time resolved velocity data from a single \change{PIV setup} with time resolved pressure measurements to estimate time resolved quantities including full velocity fluctuation fields~\citep{lee2024superresolution} and SPOD modes~\citep{zhang2020spectral}.

The rest of the paper is outlined as follows. In section \ref{sec:spod} and \ref{sec:DMD}, SPOD and DMD are formally introduced and discussed. Section \ref{sec:alg} proposes the pairwise SPOD algorithm and section \ref{sec:results} compares the results with that of standard SPOD. Finally, section \ref{sec:conclusion} has some final remarks and discusses future directions. 
% Put \label in argument of \section for cross-referencing
%\section{\label{}}

\section{Methods\label{sec:methods}}
\subsection{Spectral proper orthogonal decomposition\label{sec:spod}}
SPOD is the eigendecomposition of the cross-spectral density (CSD) tensor at discrete frequencies, producing energy-ranked, orthogonal modes~\citep{towne2018spectral,schmidt2020guide}. To compute SPOD modes, the cross-spectral density tensor is typically estimated using a Welch-like method. A long data set is separated into $N_\text{E}$  overlapping segments and a discrete Fourier transform (DFT) of each segment is taken.  Each segment is essentially considered a trial of an independent experiment, and the CSD converges as both the total data and segment size are increased.  

The CSD tensor at frequency $\omega$ is estimated as
\begin{equation}
    \mb S_\omega=\hat{\mb Q}_{\omega}\hat{\mb Q}_{\omega}^*,
\end{equation}
where $\hat{\mb Q}_{\omega}=
\left[\begin{array}{cccc}
    \hat{\vb q}^{(1)}_\omega  & \hat{\vb q}^{(2)}_\omega & \hdots & \hat{\vb q}^{(N_\text{E})}_\omega \\
\end{array}\right]$ and $\hat{\vb q}^{(j)}_\omega \in \mathbb{C}^m$ is the estimated DFT mode of the flow data for segment $j$ at frequency $\omega$, computed using a fast Fourier transform (FFT) over the segment. In the discrete setting, the flow data itself can be arbitrarily defined but is typically comprised of discrete values of the velocities, pressure, density, etc. on a grid of points.  Standard FFT implementations require sequential, time-resolved data and so SPOD has these same requirements. The SPOD modes are given through an eigendecomposition of $\mb S_\omega$,
\begin{equation}
    \mb S_\omega\mb W \boldsymbol{\Phi}_\omega = \boldsymbol{\Phi} _\omega  \boldsymbol\Lambda_\omega.
\end{equation}
Here $\mb W$ is a positive-definite Hermitian weight matrix that defines the appropriate discretized inner product. When the number of segments is much less than the snapshot dimension, $N_\text{E}\ll m$, it is computationally cheaper to estimate them using the equivalent ``snapshot'' method. The smaller $N_\text{E}\times N_\text{E}$ eigenvalue problem is,
\begin{equation}
    \hat{\mb Q}_{\omega}^* \mb W \hat{\mb Q}_{\omega} \boldsymbol{\Psi}_\omega = \boldsymbol{\Psi}_\omega  \boldsymbol\Lambda_\omega, \qquad \boldsymbol{\Phi}_\omega = \hat{\mb Q}_{\omega} \boldsymbol{\Psi}_\omega\boldsymbol \Lambda_\omega^{-1/2}.
\end{equation}
Each SPOD mode is a column of $\boldsymbol\Phi_\omega$ and the corresponding entry of $\boldsymbol \Lambda_\omega$ is the associated energy. $\boldsymbol\Psi_\omega$ are the eigenvectors of this smaller problem and are used as an intermediate variable. Ordering the modes by energy, $\boldsymbol \Lambda_\omega$, gives an optimal expansion for the Fourier mode at the given frequency. See~\citet{schmidt2020guide,heidt2024optimal} for a more detailed description of the SPOD algorithm and parameter selection.

\subsection{Dynamic mode decomposition\label{sec:DMD}}
DMD is commonly employed for state prediction, spectral analysis, and control. Originally derived using a companion matrix approach, a singular value decomposition (SVD) approach has since been more commonly implemented~\citep{Rowley2009jfm,schmid2010dynamic,chen2012variants,schmid2022dynamic}.~\citet{tu2014dmd} extended to pairwise, but not uniformly sampled sets of data, in the exact DMD formulation. Here we will focus on exact DMD and its use for state prediction and spectral analysis. Later we will leverage the state prediction ability of DMD to estimate SPOD modes using non-uniformly sampled pairwise data. 

DMD produces non-orthogonal modes with an associated complex eigenvalue that prescribes an oscillation frequency and growth/decay rate of the mode in time. Given a flow defined at time $t$ by the state vector $\vb q(t) \in \mathbb{C}^m$, DMD is the eigendecomposition of  $\mb{A}\in \mathbb{C}^{m \times m}$ given by
\begin{equation}
    \vb q(t+\Delta t)=\mb A \vb q(t),
    \label{eq:DMD}
\end{equation}
where $\mb{A}$ is found through regression using the pairwise data.  We define
\begin{equation}
    \mb X = \left[\begin{array}{cccc}
  \vb q(t_1) & \vb q(t_2) & \dots & \vb q(t_n)
    \end{array}\right] \in \mathbb C^{m\times n},
\end{equation}
and
\begin{equation}
    \mb Y = \left[\begin{array}{cccc}
        \vb q(t_1+\Delta t) & \vb q(t_2+\Delta t) & \dots & \vb q(t_n+\Delta t)
    \end{array}\right] \in \mathbb C^{m\times n},
\end{equation}
which contain the first and second snapshots of each pair, respectively. The ordering of the columns of $\mb X$ does not matter. However, the column-wise correspondence between $\mb X$ and $\mb Y$, shifted by a constant $\Delta t$, is key and must be preserved. \change{Finally, $\mb A$, as defined in equation \ref{eq:DMD}, can be approximated as
\begin{equation}
    \mb A = \underset{\mb A}{\text{argmin}}\norm{\mb Y -\mb A \mb X}_F^2,
    \label{eq:Adef}
\end{equation}
where $\norm{\cdot}_F$ denotes the Frobenius norm.} The DMD is the eigendecomposition,
\begin{equation}
    \mb A\vb \varphi=\vb \varphi\lambda,
\end{equation}
where the eigenvector $\vb \varphi$ is the spatial DMD mode with temporal dynamics associated with the corresponding eigenvalue, $\lambda$.

In many complex flow applications, $m\gg n$ and thus the pseudo-inverse is ill-conditioned. \change{By using the SVD, we can write $\mb X= \mb U \mb S \mb  V^* $; the matrices $\mb U$ and $\mb V$ are unitary and $\mb S$ is diagonal.} Under the approximation that the flow lies in a lower dimensional manifold, the problem can be regularized by exploiting the POD modes of $\mb X$, which are the columns of $\mb U$, and which provide the best (Frobenius norm) approximation of $\mb X$. 
Using this $r$-truncated representation, $\mb X\approx\mb U_r \mb S_r \mb  V_r^*$, we can define $\mb{\Tilde{X}}=\mb U_r^*\mb X$, $\mb{\Tilde{Y}}=\mb U_r^*\mb Y$, and $\mb{\tilde{A}}=\mb U_r^*\mb A \mb U_r$ to formulate a much smaller problem,
\change{
\begin{equation}
    \mb{\Tilde A} = \underset{\mb{\Tilde A}}{\text{argmin}}\norm{\mb{\Tilde Y} -\mb{\Tilde A} \mb{\Tilde X}}_F^2=\mb{\Tilde Y}\mb {\Tilde X}^+\in \mathbb C^{r \times r},
\end{equation}
where the $\mb {\Tilde X}^+$ denotes the pseudoinverse since $r\leq n$. Using the SVD of $\mb {X}$ we can compute $\mb{\Tilde A}$ as}
\begin{equation}
    \mb{\tilde A}=\mb U_r^* \mb{Y}\mb{V}_r\mb{S}_r^{-1} \in \mathbb C^{r \times r}.
\end{equation}
The eigendecomposition is then given by
\begin{equation}
    \mb{\tilde A}\vb w=\vb w \lambda.
\end{equation}
For each non-zero $\lambda$, the corresponding high-dimensional DMD mode can be computed as
\begin{equation}
    \vb {\varphi} = \mb{Y}\mb{V}_r\mb{S}_r^{-1} \vb w.
\end{equation}
Each mode has an associated oscillation and growth/decay rate of $ \arg(\lambda)/(2 \pi \Delta t)$ and $|\lambda|/(2 \pi \Delta t)$ respectively. Computing the eigendecomposition of $\mb{\Tilde A}$ is much more computationally tractable due to the dimension reduction, $r\leq n \ll m$. In addition, ~\citet{tu2014dmd} showed that the set of eigenpairs given by $(\lambda_j,\vb \varphi_j)$ are exactly the non-zero eigenpairs of $\mb A$. A reduced order model for the time evolution of the system is given by
\begin{equation}
    \vb q(t)=\sum_{j=1}^r \vb \varphi_j \me^{\log(\lambda_j)\, t/\Delta t }b_j,
\end{equation}
where $b_j$ is the initial amplitude of $\vb \varphi_j$ at $t=0$. 

DMD may produce decaying and growing modes, despite statistical stationarity, due to noise. \citet{bagheri2014effects} showed that the DMD spectrum of containing limit cycles becomes damped \change{in the presence of process noise or phase drift.} This means exact DMD estimates decaying eigenvalues for statistically stationary flows. These modes lose energy as we extrapolate farther from the initial condition. Several formulations of DMD have been developed to overcome the effects of \change{both process noise and sensor noise}~\citep{Jovanovic2014pof,hemati2017biasing,askham2018variable,baddoo2023physics}. One such method, forward-backward DMD (fbDMD)~\citep{dawson2016characterizing}, estimates a new matrix $\mb B\in \mathbb{C}^{m \times m}$ that approximately estimates the state prior,
\begin{equation}
    \vb q (t)=\mb B \vb q(t+\Delta t).
\end{equation}
This matrix is computed analogously to $\mb A$, but with $\mb X$ and $\mb Y$ swapped. The fbDMD is then the eigendecomposition of $(\mb A \mb B^{-1})^{1/2}$, denoting the matrix square root. Despite this, DMD is still limited in how far it can extrapolate from the initial condition; however, it gives us a way to estimate the flow at any time given only pairwise data.

\subsection{Proposed pairwise SPOD algorithm\label{sec:alg}}
\begin{figure*}[tb]
    % \centering 
    \includegraphics[scale=0.99]{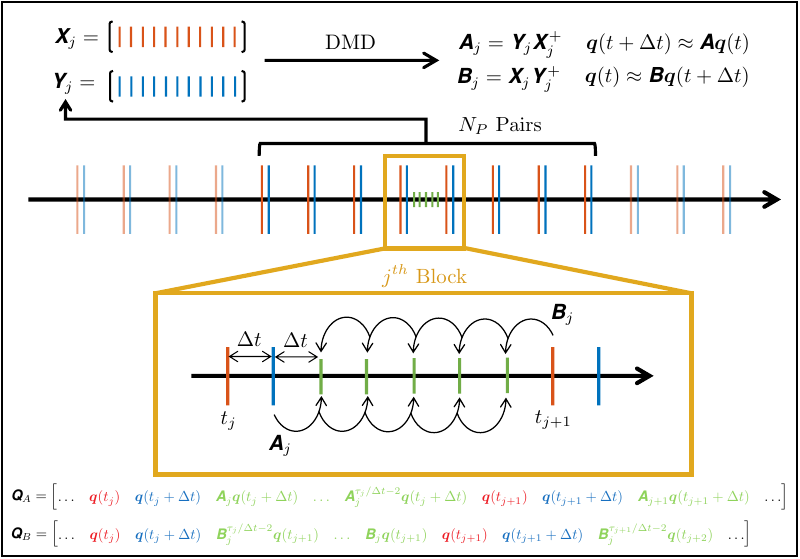}
    \caption{Schematic of estimating one block of missing data as part of pairwise SPOD. The given pairwise non-uniformly sampled data is depicted by red and blue representing the first and second snapshot of each pair, respectively. The snapshots being estimated are shown in green. To estimate the missing data, the $N_P$ closest pairs are used to create an $\mb A$ and $\mb B$ matrix. Starting from the second snapshot of pair $j$, DMD is used to estimate the snapshots missing in the block. Once $t_{j+1}$ is reached, the process is repeated by shifting the block over one pair and estimating that block's missing snapshots. $\mb B$ matrix is analogously propagated from right to left.
    \label{fig:Alg}}
\end{figure*}
The proposed algorithm, pairwise SPOD, is now introduced. Given pairwise data, the algorithm estimates segments of sequential data using several local (in time) DMD realizations. After this step, the resulting sequential data can be processed with the standard SPOD algorithm described in section \ref{sec:spod}.

Pairwise data is given as follows. The $j^{th}$ pair is taken at time $t_j$ and $t_j+\Delta t$, where $\Delta t$ is fixed and satisfies the Nyquist criterion. We define $\tau_j=t_{j+1}-t_j$ as the camera delay time, which is the time a camera has to recover between pair $j$ and $j+1$. We restrict ourselves to the case where there is an integer number of time steps between pairs, $\tau_j/\Delta t\in\mathbb{Z}^+$. The $\tau_j/\Delta t-2$ missing snapshots between the pairs are estimated using DMD, as shown in figure \ref{fig:Alg}. Each data set has $\tau_{\min}$ and $\tau_{\max}$ defining the minimum and maximum delay time between pairs, respectively.
We let 
\[
\tau_j\sim\mathcal{U}\left( \tau_{\min},\, \tau_{\max}\right)
\]
vary uniformly between these bounds to emulate the random sampling procedure used in~\citet{tu2014spectral}. We emphasize that $\tau$, the gap time, varies between pairs but the time within a pair, $\Delta t$, is constant. This strategy decreases the chance of peaks at $f=1/\tau$ due to a constant delay time. Additionally, it allows us to capture all phases of the flow. \change{The simplification to integer number of time steps between pairs is made here to maximize the number of actual snapshots in the reconstruction and is not required for exact DMD. If $\Delta t$ is not constant, another version of DMD such as $\theta-$DMD~\citep{li2022dynamic} or optimal DMD~\citep{askham2018variable} may be used; however, this is not explored here. }

To avoid large discontinuities at the segment ends, and as an attempt to combat the non-physically damped eigenvalues of the DMD matrix, two estimates of the data will be produced - one marching forward and one marching backward in time. The final estimate used in SPOD is taken as a weighted linear combination of the forward and backward estimates. This slightly differs from, but is inspired by, fbDMD.

To compute the forward estimation as sketched in figure \ref{fig:Alg}, for each stretch of missing data, the $N_P$
closest pairs to the missing data (in time) are used to create a local $\mb A$. A spatially global, segment-wise, POD basis of $r$ truncated modes is used. The second snapshot of the closest pair to the left is used as the initial condition to estimate the missing snapshots by continuously applying equation \ref{eq:Adef} until the next pair is reached. The next section (Block $\change{j}+1$) is estimated similarly; however, a new, local, $\mb A$ matrix is created with the new $N_P$ closest pairs, like a sliding window. This continues until the end of the pairs is reached. This sequential estimate is a combination of known snapshots (the given pairs) and estimated snapshots (from DMD).

The backward estimate is computed analogously; however, the matrix $\mb B$ is computed as $\mb B=\mb X \mb Y^+$ which approximately satisfies $\vb q(t)=\mb B \vb q(t+\Delta t)$. The march is started from the first snapshot of the closest pair on the right rather than the left.

The final sequential estimate is computed as
\begin{equation}
    \mb Q(t) = \alpha(t)\mb Q_{\mb A} + (1-\alpha(t))\mb Q_{\mb B},
\end{equation}
where $\alpha$ decreases linearly from 1 to 0 between $t_j$ and $t_{j+1}$ for $j=1,2,\hdots, N$, and $\mb Q_{\mb A}$ and $\mb Q_{\mb B}$ are the estimates from the forward and backward march, respectively. A summary of the parameters needed to specify the algorithm is given in table \ref{tab:parameters} \change{and pseudocode is shown in algorithm \ref{alg:1}}.

\change{The application of DMD is ideal for SPOD estimation due to the connection between SPOD and DMD~\citep{towne2018spectral}. In the limit that the pairwise data is actually sampled at a constant rate (ie. $\tau=2\Delta t)$, the DMD modes are DFT modes which is equivalent to standard SPOD. There are several other methods for snapshot reconstruction such as Kalman smoothing used in dual-PIV in~\citet{kaneko2024dmd} however we use equation \ref{eq:Adef} because the connection between SPOD and DMD~\citep{towne2018spectral}}.

\begin{algorithmfloat}[\change{Pseudocode for Pairwise SPOD}]
\label{alg:1}
\begin{algorithmic}
\State Input: $\mb X \in \mathbb{C}^{m\times n},\mb Y\in \mathbb{C}^{m\times n},\vb \tau\in \mathbb{R}^n, \Delta t, N_{\textrm{P}},r $ 
\For{$j=N_{\textrm{P}}/2; j<n-N_{\textrm{P}}/2$}

 $\mb X_j=\begin{bmatrix}
     \mb x_{j+1-N_{\textrm{P}}/2 } & \mb x_{j+2-N_{\textrm{P}}/2} & \hdots & \mb x_{j} & \mb x_{j+1}& \cdots \mb x_{j+N_{\textrm{P}}/2}
 \end{bmatrix}$ \Comment{$\mb x_j=\vb q(t_j)$}
 
  $\mb Y_j=\begin{bmatrix}
     \mb y_{j+1-N_{\textrm{P}}/2 } & \mb y_{j+1-N_{\textrm{P}}/2} & \hdots & \mb y_{j} & \mb y_{j+1}& \cdots \mb y_{j+N_{\textrm{P}}/2}
 \end{bmatrix}$\Comment{$\mb y_j=\vb q(t_j+\Delta t)$}

 $\mb A_j=\textrm{ExactDMD}(\mb X_j, \mb Y_j, r)$
 
 $\mb B_j=\textrm{ExactDMD}(\mb Y_j, \mb X_j, r)$
 
 $\mb Q.\textrm{append}(\mb y_j)$ \Comment{Second snapshot of current pair}
 
 \For{$k=1; k<\tau_{j}/\Delta t-2$}

    % \hspace{5mm}$\mb q = \frac 1 2 (\mb A_j^k\mb y_j + \mb B_j^{(\tau_{j}/\Delta t-k-1)}\mb x_{j+1})$ \Comment{Estimate kth snapshot}
    \hspace{5mm} $\mb q_A=\mb A_j^k\mb y_j$\Comment{Forward estimation}

    \hspace{5mm} $\mb q_B=\mb B_j^{(\tau_{j}/\Delta t-k-1)}\mb x_{j+1}$\Comment{Backward estimation}

    \hspace{5mm} $\alpha=1-\frac{1}{\tau_{j}/\Delta t-1}k $
    .pdf
    \hspace{5mm}$\mb Q.\textrm{append}(\alpha \mb q_A + (1-\alpha)\mb q_B)$ \Comment{Combined estimation}
 
 \EndFor
 
  $\mb Q.\textrm{append}(\mb x_{j+1})$ \Comment{First snapshot of next pair}
 
\EndFor

$[\boldsymbol\Lambda, \boldsymbol{\Psi},\omega]=\textrm{SPOD}(\mb Q)$
\State Output: $\boldsymbol\Lambda, \boldsymbol{\Psi},\omega$
\end{algorithmic}
\end{algorithmfloat}

\begin{table*}[tb]
    % \centering
    % \renewcommand{\arraystretch}{1.5}
    \begin{ruledtabular}
    \begin{tabular}{ll}
        $N_P$ & Number of pairs used to construct DMD matrix $\mb A$ or $\mb B$\\
        $r$ &  Truncation of POD basis, $r \leq N_P$\\
        $\tau_{\min}$ & Minimum time between the snapshots of adjacent pairs\\
        $\tau_{\max}$ & Maximum time between the snapshots of adjacent pairs\\
        $\overline\tau$ & Average time between adjacent pairs, $(\tau_{\min}+\tau_{
        \max})/2$\\
        $N_\text{fft}$ & SPOD block length. Number of snapshots in each FFT realization\\
        $N_\text{ovlp}$ & Number of overlapping snapshots in adjacent SPOD blocks
    \end{tabular}
    \end{ruledtabular}
    \caption{Pairwise SPOD parameters. For more on SPOD parameter selection, see~\citet{schmidt2020guide}.
    \label{tab:parameters}}
\end{table*}

\section{Results\label{sec:results}}
The following section shows the results of applying pairwise SPOD to two data sets: a test problem using the linearized Ginzburg-Landau equations and a large eddy simulation (LES) of a turbulent jet. In both cases, the pairwise SPOD results are compared to the standard uniformly-sampled SPOD modes. 
\begin{figure}
    % \centering
    \begin{subfigure}[b]{0.47\columnwidth}
        \includegraphics{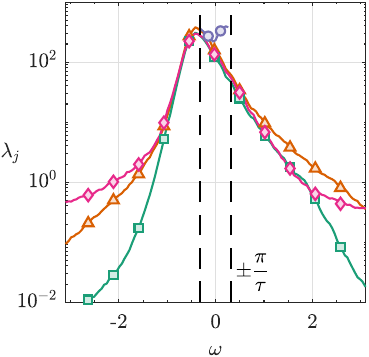}
        \caption{Mode 1}
    \end{subfigure}
    \begin{subfigure}[b]{0.47\columnwidth}
        \includegraphics{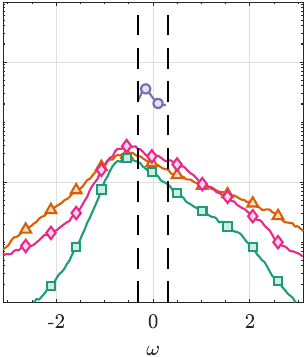}
        \caption{Mode 3}
    \end{subfigure}
    \caption[]{Comparison of (a) dominant and (b) second sub-dominant SPOD mode spectrum. (Orange, \markertri) Full Data, (Green, \markersquare) pairwise SPOD with $\overline\tau=20\Delta t$, \change{(Pink,\markerdiamond) pairwise SPOD using only forward estimation,} (Purple, \markercircle) Full data downsampled by 20. Data markers are used to improve readability and do not mark every data point.\label{fig:GLSpec}}
    
\end{figure}
\subsection{Linearized Complex Ginzburg-Landau Equation}

The method will first be tested on a data set modeled by the linearized complex Ginzburg-Landau equation. The equation takes the form 
\begin{equation}
    \frac{\partial \vb q}{\partial t}=\mathcal{A}\vb q + u(x,t),
\end{equation}
where
\begin{subequations}
    \begin{align}
    \mathcal{A}&=-\nu\frac{\partial}{\partial x}+\gamma\frac{\partial^2}{\partial x^2}+\mu(x),\\
% \end{equation}
% \begin{equation}
    \mu(x)&=(\mu_0-c_{\mu}^2)+\frac{\mu_2}{2}x^2,
\end{align}
\end{subequations}
and $u(x,t)$ is an applied external forcing. The parameters and spatial dependence used are taken to be the same as in~\citet{towne2018spectral}. This choice results in a globally stable system. The system is solved using a pseudo-spectral method as in~\citet{bagheri2009input,towne2018spectral}. The results are interpolated onto a uniform 220-node grid where $-85\leq x\leq 85$, and therefore $\mathbf{W}=\Delta x \mathbf{I}$. 

For the following results, the PDE is forced with bandlimited $(0.6 f_s)$ white noise forcing with a variance of 1. Additionally, the forcing is limited to a portion of the spatial domain; the variance decreases exponentially at the boundaries. This forcing is the same as the white noise forcing case in~\citet{towne2018spectral}. All pairwise SPOD results are computed with the parameters $N_P=200$, $r=40$ and $\Delta t=0.5$.  That is, in each block, 200 data pairs are used to estimate 40 DMD modes that best represent the dynamics of the first 40 POD modes of $\mb X_j\in\mathbb{C}^{220\times N_P}$. This truncation number was chosen from a few tests to maximize the energy captured while ensuring a well-conditioned pseudo-inverse.

We first compare the pairwise SPOD results for a single $\overline\tau$. The following pairwise results are with parameters $\tau_{\min}=10\Delta t$ and $\tau_{\max}=30\Delta t$, which leads to $\overline{\tau}=20\Delta t$. Therefore, between pairs, an average of 18 snapshots need to be estimated. To test the anti-aliasing effects, a sequential SPOD is also computed on the full data downsampled by $\overline\tau/\Delta t$, where $\overline\tau=20 \Delta t$. \change{Additionally, pairwise SPOD using both the forward and backward DMD operators is compared to using the forward prediction alone.}

Figure \ref{fig:GLSpec} compares the first and third SPOD eigenvalues for the full data, pairwise ($\overline\tau=20 \Delta t$) and downsampled (by 20). The data is complex, which leads to a 2-sided spectrum. The final pairwise march produces estimates of 96,000 sequential snapshots. 96,000 snapshots of the full and pairwise estimates are used with a Hamming window of length $N_\text{fft}=240$ for our SPOD estimation (see~\citet{schmidt2020guide} for more on SPOD parameter selection). The spectra clearly show the power of pairwise SPOD for dealiasing the SPOD spectrum within the range of a downsampled frequency. Additionally, pairwise SPOD gives good estimates for the energy outside of this range, particularly for $-1\leq\omega\leq1$. The overall decrease in energy is due to the artificially damped eigenvalues at high frequencies that DMD produces with statistically stationary data in the presence of \change{process} noise~\citep{bagheri2014effects,hirsh2020centering,scherl2020robust}. This effect is stronger for the sub-dominant SPOD modes that may be less converged. While other versions of DMD have been developed to combat this effect~\citep{chen2012variants,dawson2016characterizing,baddoo2023physics}, we found these versions did not improve these results when combined with the proposed pairwise SPOD algorithm.  \change{We also note that the forward-only estimation has a much more energetic spectrum than the forward-backward.}

\begin{figure}
    % \centering
    \begin{subfigure}[b]{0.4\columnwidth}
        % \centering
        \includegraphics[]{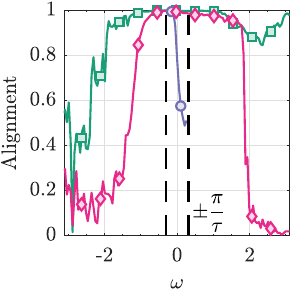}
        \caption{Mode 1}
    \end{subfigure}
    \begin{subfigure}[b]{0.4\columnwidth}
        % \centering
        \includegraphics[]{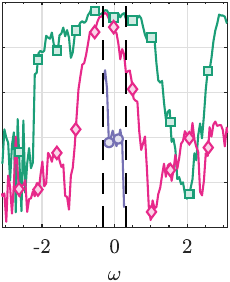}
        \caption{Mode 3}
    \end{subfigure}
    \caption[]{Alignment, $\langle\boldsymbol{\Phi}_{est}, \boldsymbol{\Phi}_{full}\rangle$, between full-data sequential SPOD and three different SPOD estimates: (Green, \markersquare) pairwise SPOD with $\overline\tau=20\Delta t$ , \change{(Pink, \markerdiamond) pairwise SPOD using only forward prediction,} and (Purple, \markercircle) downsampled sequential data for the (a) dominant and (b) second sub-dominant modes.
    \label{fig:GLAlign}}
\end{figure}

We use alignment as a quantitative measure of how accurately pairwise SPOD estimates mode shapes. Alignment is defined as the weighted inner product of the estimated mode shape with the full data SPOD mode. Since the modes form an orthonormal set, an alignment of 1 means the modes are the same and 0 means they are orthogonal.

Figure \ref{fig:GLAlign} shows the alignment of the pairwise and downsampled estimates with the fully resolved modes. The mode 1 alignment shows that the \change{forward-backward} pairwise data can give accurate estimates of mode shape, particularly for $-2\leq\omega\leq2$. As expected, the (far less energetic) mode 3 estimates are globally less aligned than the mode 1 estimates. \change{The forward only pairwise estimation is less aligned then the foward-backward estimate at all frequencies, particularly for mode 3.} By contrast, the downsampled estimate is severely aliased and does not produce reliable mode shape estimates at any frequency. 

\change{We now note the differences between using just a forward estimator rather than the combined estimator previously discussed. In figure \ref{fig:GLSpec}, the forward estimate gives a more accurate prediction of the PSD, however in figure \ref{fig:GLAlign}, the alignment clearly suffers particularly at higher frequencies. The combined estimate is more damped than just the forward estimate but produces more accurate mode shapes. For the rest of the paper we will use only the combined estimation for pairwise SPOD.}

\begin{figure}
    % \centering
    \begin{subfigure}[b]{0.47\columnwidth}
        \caption{Mode 1 full data}
        % \centering
        \includegraphics[]{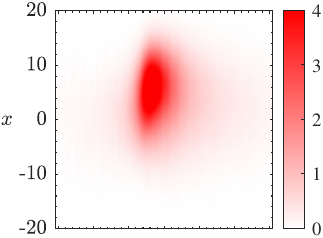}
    \end{subfigure}
    \begin{subfigure}[b]{0.47\columnwidth}
        \caption{Mode 3 full data}
        % \centering
        \includegraphics[]{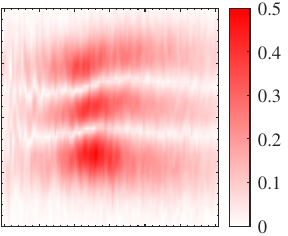}
    \end{subfigure}
    \begin{subfigure}[b]{0.47\columnwidth}
        \caption{Mode 1 $\overline\tau=20\Delta t$}
        % \centering
        \includegraphics[]{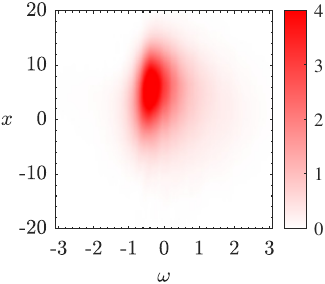}
    \end{subfigure}
    \begin{subfigure}[b]{0.47\columnwidth}
        \caption{Mode 3 $\overline\tau=20\Delta t$}
        % \centering
        \includegraphics[]{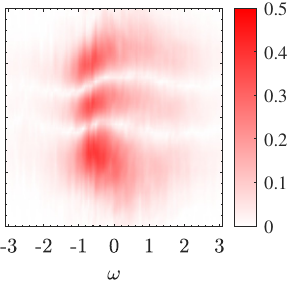}
    \end{subfigure}
    \caption{Weighted SPOD mode shapes, $\sqrt{\lambda_j(\omega)}\lvert \vb \psi_j(x,\omega) \rvert$. (a,b) Full data and (c,d) $\overline\tau=20\Delta t$ modes shapes for the (a,c) dominant and (b,d) second sub-dominant SPOD mode.\label{fig:GLModes}
    }
    
\end{figure}

Figure \ref{fig:GLModes} shows the weighted mode shapes for the first and third modes at each frequency. The structure of mode 1 is captured well by pairwise SPOD.  The structure of mode 3 shows the effect of the decreasing energy at high frequencies in the pairwise SPOD; however, the distribution of energy in the $x$-direction is well captured at the lower frequencies by the pairwise data. 

\begin{figure}
        \begin{subfigure}[b]{0.6\columnwidth}
        % \centering
        \includegraphics[]{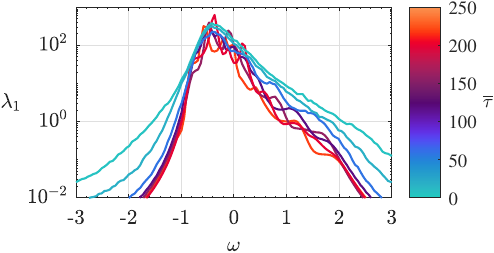}
        \caption{}
    \end{subfigure}
    \begin{subfigure}[b]{0.35\columnwidth}
        % \centering
        \includegraphics[]{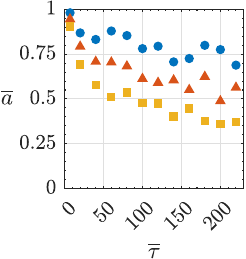}
        \caption{}
    \end{subfigure}
    \caption[]{Pairwise SPOD as a function of $\overline\tau$. \change{For $\overline\tau=7$, $\tau_{\max}-\tau_{\min}=6\Delta t$, $\tau_{\max}-\tau_{\min}=20\Delta t$ for all other $\overline\tau$.} (a) Mode 1 spectrum. (b) Average alignment for $\omega\in[-3,\, 3]$ of (Blue, \markercircle) mode 1, (Red, \markertri) mode 2 and (Yellow, \markersquare) mode 3.
    \label{fig:GLTauVary}}
\end{figure}

We will now investigate the effect of changing the average camera delay time. Figure \ref{fig:GLTauVary} shows how the dominant SPOD spectrum and average alignment of the first 3 mode shape estimates are affected by an increase in $\overline\tau$. \change{Here we use the same SPOD parameters as before but the total number of snapshots estimated in each case is 50,000.} We define the average alignment $\overline a$ as a normalized integration of the alignment over $-3\leq\omega\leq3$, the forcing frequency range,
\begin{equation}
    \overline a_j = \frac 1 6 \int_{-3}^{3} a_j(\omega)\, d\omega,
\end{equation}
where $a_j(\omega)$ is the alignment of the estimated SPOD mode $j$ with the full data mode $j$ at frequency $\omega$.
As $\overline\tau$ increases, the spectrum is progressively attenuated up to approximately $\overline\tau=100\Delta t$, after which spurious peaks start to occur in the spectrum. This is likely the result of slowly growing DMD modes with small initial amplitude becoming dominant.  Further increase of $\overline\tau$  (not shown) yields very poor results. The average alignment falls off from unity as $\overline\tau$ is increased, as expected, but stabilizes after $\overline\tau\approx50\Delta t$.

The Ginzburg-Landau data shows the efficacy of pairwise SPOD for estimating SPOD spectra and modes.  Up to the Nyquist frequency associated with the gap, the results are in good quantitative agreement with sequential POD (dealiasing) and also remain qualitatively correct up to about twice the Nyquist frequency.  

In this example, there was essentially no limitation on the number of pairs available as the Ginzburg-Landau system is relatively small. In the next section, we provide a more realistic example using turbulent flow data.

\subsection{Turbulent Jet}

The method is now tested with high-fidelity simulation data for a Mach $0.4$, isothermal turbulent jet at a Reynolds number of $450\times10^3$. The data was computed using the Charles solver by Cascade Technologies and was previously experimentally validated~\citep{bres2017unstructured,bres2018importance}. Comparisons of sequential SPOD and DMD modes for this data were previously investigated in~\citet{towne2018spectral}. The jet is first decomposed into Fourier modes in the azimuthal direction. For brevity, we focus on the axisymmetric component. The data set consists of 20,000 sequential snapshots sampled every $\Delta t =0.2$ acoustic time units. The data is fully time-resolved and consists of the density, 3-component velocity, and temperature fields. Thus, it is possible to generate subsampled data sets, representative of experimental measurements, and compare the SPOD predictions with the ground truth from the fully resolved data.  The state vector is in the form 
$\vb q=\begin{bmatrix}
    \rho & u_x & u_r & u_{\theta} & T
\end{bmatrix}^\top$, where each term is the fluctuation from the long time averaged quantity at that field point. We adopt the Chu compressible energy norm~\citep{chu1965energy} and weighting matrix used in previous analysis~\citep{towne2018spectral}. The computational domain extends to 6 and 30 times the jet diameter in the radial and streamwise directions, respectively. Because the fully time-resolved data is available, the full data SPOD modes can be compared to the results of pairwise SPOD.

\begin{figure}
    % \centering
    \begin{subfigure}[b]{0.48\columnwidth}
        % \centering
        \includegraphics[]{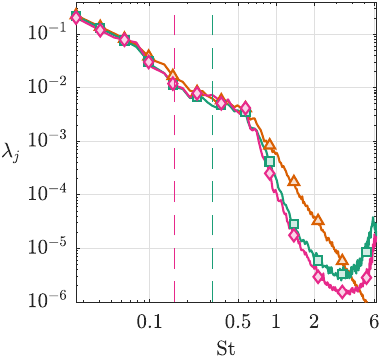}
        \caption{Mode 1}
    \end{subfigure}
    \begin{subfigure}[b]{0.48\columnwidth}
        % \centering
        \includegraphics[]{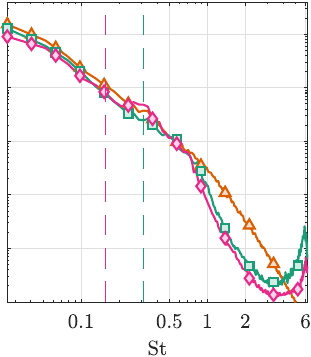}
        \caption{Mode 2}
    \end{subfigure}
    \caption[]{Comparison of (a) first and (b) second SPOD eigenvalues of the turbulent jet data. (Orange, \markertri) Full Data, (Green, \markersquare) $\overline\tau=20\Delta t$, (Pink, \markerdiamond) $\overline\tau=40\Delta t$. Dashed lines correspond to the gap Nyquist frequency $\pi / \overline\tau$.  Frequencies are reported in terms of Strouhal number, $St=\frac{fD}{U_j}$.
    \label{fig:JetSpec}}
\end{figure}

In figure~\ref{fig:JetSpec}, two different pairwise data sets are shown. The first data set, denoted $\bar{\tau}=20$, has a camera gap time between 10 and 30 snapshots; the second case has $\bar{\tau}=40$. Both pairwise SPOD estimates use $N_P=r=300$. For all cases, SPOD is computed using a Hamming window of $N_\text{fft}=480$ with 50\% overlap as suggested in~\citet{heidt2024optimal}.

The sequential SPOD modes are computed using 15,000 snapshots while only 2,000 actual snapshots (1,000 pairs) and 1,000 actual snapshots (500 pairs) are used for the $\bar{\tau}=20\Delta t$, and $\bar{\tau}=40\Delta t$ estimates, respectively. The rest of the pairwise snapshots are estimates from the DMD march. This means that the $\bar{\tau}=40\Delta t$ case is using 94\% less  data than the full data case, which serves as the basis for estimation. 

The pairwise SPOD eigenvalues of the first two modes are compared to the sequential SPOD spectrum in figure~\ref{fig:JetSpec}. Both the $\bar{\tau}=20\Delta t$ and $\bar{\tau}=40\Delta t$ data sets agree well with the full data for $St\lesssim0.7$. The decrease in energy at higher frequencies can again be attributed to the damped eigenvalues in the DMD matrix. This results in a decrease in energy with each application of $\mb A$, which is consistent with $\overline{\tau}=40$ having less energy than $\overline{\tau}=20$.  Physically, the range $\St\lesssim0.7$ includes the important, largest-scale coherent structures in this jet, and this result is thus of immediate practical benefit as camera speeds can be reduced by a factor of 40.
\begin{figure}
    % \centering
    % \captionsetup{justification=centering}
    \begin{subfigure}[b]{0.4\columnwidth}
        % \centering
        \includegraphics[]{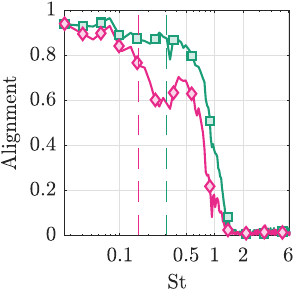}
        \caption{Mode 1}
    \end{subfigure}
    \begin{subfigure}[b]{0.4\columnwidth}
        % \centering
        \includegraphics[]{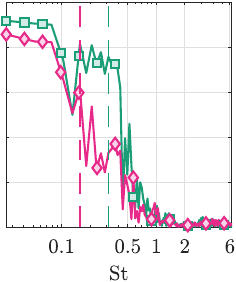}
        \caption{Mode 2}
    \end{subfigure}
    \caption[]{Comparison of (a) first and (b) second SPOD mode alignment of turbulent jet with full data modes. (Green, \markersquare) $\overline\tau=20\Delta t$, (Pink, \markerdiamond) $\overline\tau=40\Delta t$. Dashed lines correspond to the Nyquist frequency of the full data downsampled by $\overline\tau/\Delta t$.
    \label{fig:JetAlign}}
\end{figure}
Figure~\ref{fig:JetAlign} shows the alignment between pairwise and  sequential SPOD modes for modes 1 and 2 for both gap times.  Both pairwise cases have reasonable alignment with the actual modes up to $St\approx1$. Increasing $\overline{\tau}$ results in a global decrease in alignment. This agrees with what was seen in the Ginzburg-Landau experiments. 

Additionally, figure~\ref{fig:JetModes} compares the mode shapes at several Strouhal numbers, $St=0.1,\, 0.6,\, 1.2$. The mode shape at $St=0.6$ agrees quantitatively in both cases.  For both $\bar{\tau}=20\Delta t$ and $40\Delta t$, the frequencies $St=0.6$ and $St=1.2$ are beyond the gap Nyquist frequency. At $\St=1.2$, both pairwise estimates have very low alignment with the full data; however, they both correctly estimate a decrease in wave number. The alignment of $\tau=40$ may be improved if more data pairs were available, as we believe that increasing $N_P$ will give us better estimates at high frequencies.

\begin{figure}
    % \centering
    \begin{subfigure}[b]{0.48\columnwidth}
    % \centering
    \includegraphics{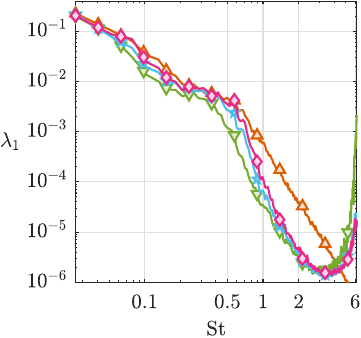}
    \caption{Mode 1 Spectrum}
    \end{subfigure}
    \begin{subfigure}[b]{0.48\columnwidth}
    % \centering
    \includegraphics{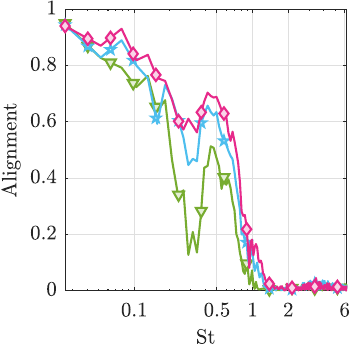}
    \caption{Mode 1 Alignment}
    \end{subfigure}
    \caption[]{Comparison of (a) mode 1 spectrum and (b) mode 1 alignment for (Orange,\markertri) Full data SPOD modes, and pairwise SPOD modes with parameters: (Pink,\markerdiamond) $\overline\tau=40\Delta t$, $N_P=300$,(Blue,\markerstar) $\overline\tau=40\Delta t$, $N_P=200$,(Green,\markertridown) $\overline\tau=40\Delta t$, $N_P=100$.
    \label{fig:JetNP}}
\end{figure}

Figure~\ref{fig:JetNP} shows the dominant mode spectrum and alignment for $\overline\tau=40$ for $N_P=\,300,\,200,\, 100$. As $N_P$ is increased, the estimate recovers more energy in the flow at all frequencies. The alignment is slightly improved at higher frequencies by increasing $N_P$. Increasing $N_P$ further may result in even better pairwise estimations; however, we are limited in our testing abilities with this jet data set because of the number of snapshots available.

\newcommand{\MM}{0.25}
\begin{figure*}
    % \centering
    \includegraphics[scale=\MM]{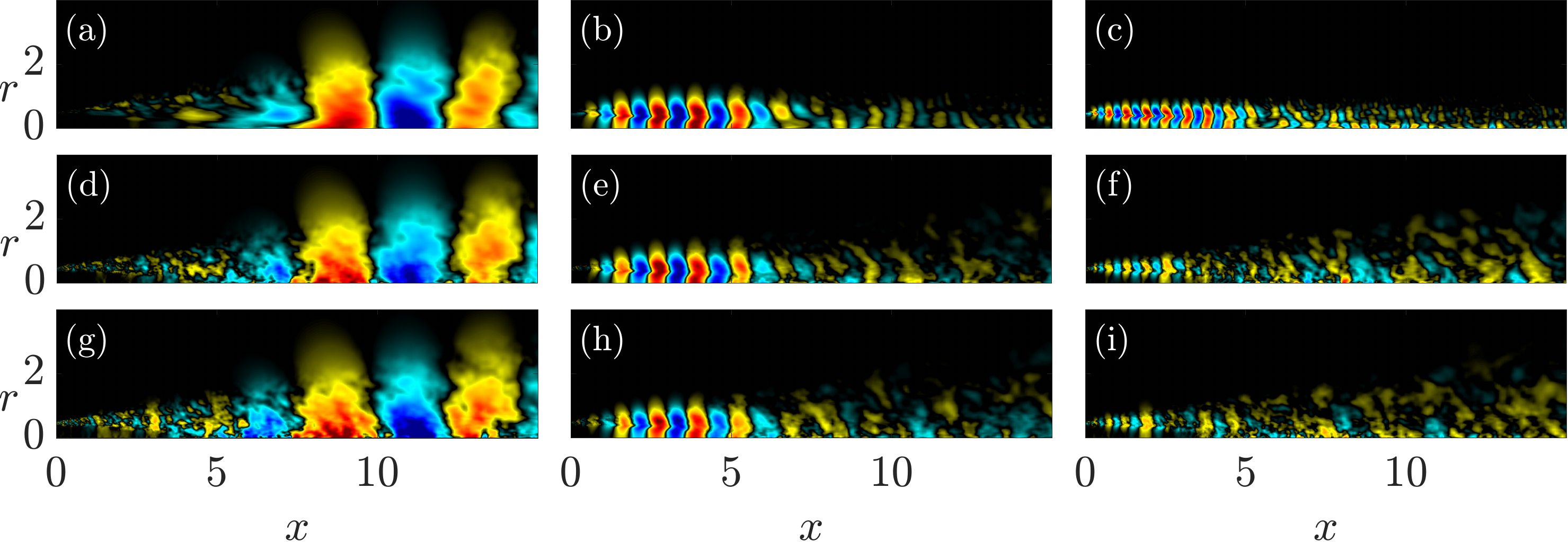}
    \caption{Real component of pressure of dominant SPOD mode shape for (a-c) Full data, (d-f) $\overline\tau=20\Delta t$ and (g-i) $\overline\tau=40\Delta t$ at frequency (a,d,g) $\St=0.1$, (b,e,h) $\St=0.6$, (c,f,i) $\St=1.2$. Pressure is computed using a linearized equation of state.
    \label{fig:JetModes}}
\end{figure*}

\section{Conclusion\label{sec:conclusion}}

We have developed an algorithm to estimate the SPOD spectrum for sub-Nyquist data in the form of data pairs separated by long gaps.  The algorithm first estimates uniformly sampled data using exact DMD, after which the standard SPOD algorithm can be applied.  We tested the algorithm on two data sets: linearized complex Ginzburg-Landau and an LES of a turbulent jet. 
The jet results showed that the dominant coherent structures could be resolved despite an acquisition rate 40 times slower than time-resolved data. 

While the algorithm is largely motivated by camera-speed limitations in PIV, the work has other implications. Even for slower flows where time-resolved PIV is possible with a single camera, \change{reducing the acquisition rate} allows increased resolution. In addition, for very large simulations, saving sequential data can be prohibitive -- pairwise SPOD can greatly reduce the required disk space. In the case of the turbulent jet, the total number of "saved," actual snapshots used in $\overline\tau=40\Delta t$ was 1014, while the full data used 15,000 snapshots. This is a 93\% decrease in the amount of data stored during the simulation, independent of $N_P$.

To extend pairwise SPOD to an even more general domain, it may be possible to use optimal DMD~\citep{askham2018variable} or $\theta$-DMD~\citep{li2022dynamic}, which are formulations that do not require pairwise data.  In addition,~\citet{asztalos2023galerkin} estimated SPOD modes using randomly sampled data by estimating missing snapshots using a physics-based POD-Galerkin projection model. Combining the physics-based model proposed in~\citet{asztalos2023galerkin} and the purely data-driven model proposed here may be possible through data assimilation techniques. Additionally, extending pairwise SPOD to related methods including bispectral mode decomposition (BMD)~\citep{schmidt2020bispectral} could be explored.

Finally, we demonstrated that utilizing pairwise data allows us to extract these coherent structures and understand their temporal dynamics by using DMD as a tool for interpolation. However, the resulting SPOD spectrum is a function of the different DMD operators themselves. This suggests there may be a way to extract the SPOD spectrum from the DMD spectrum directly as a replacement for a snapshot estimation marching scheme proposed here.

\begin{acknowledgments}
% put your acknowledgments here.
This work was supported by Boeing under award CT-BA-GTA-1. C.C. would also like to acknowledge funding from the National Science Foundation GRFP under DGE‐1745301.

\end{acknowledgments}

% Create the reference section using BibTeX:
\bibliography{refs}

%apsrev4-2.bst 2019-01-14 (MD) hand-edited version of apsrev4-1.bst
%Control: key (0)
%Control: author (8) initials jnrlst
%Control: editor formatted (1) identically to author
%Control: production of article title (0) allowed
%Control: page (0) single
%Control: year (1) truncated
%Control: production of eprint (0) enabled
\begin{thebibliography}{32}%
\makeatletter
\providecommand \@ifxundefined [1]{%
 \@ifx{#1\undefined}
}%
\providecommand \@ifnum [1]{%
 \ifnum #1\expandafter \@firstoftwo
 \else \expandafter \@secondoftwo
 \fi
}%
\providecommand \@ifx [1]{%
 \ifx #1\expandafter \@firstoftwo
 \else \expandafter \@secondoftwo
 \fi
}%
\providecommand \natexlab [1]{#1}%
\providecommand \enquote  [1]{``#1''}%
\providecommand \bibnamefont  [1]{#1}%
\providecommand \bibfnamefont [1]{#1}%
\providecommand \citenamefont [1]{#1}%
\providecommand \href@noop [0]{\@secondoftwo}%
\providecommand \href [0]{\begingroup \@sanitize@url \@href}%
\providecommand \@href[1]{\@@startlink{#1}\@@href}%
\providecommand \@@href[1]{\endgroup#1\@@endlink}%
\providecommand \@sanitize@url [0]{\catcode `\\12\catcode `\$12\catcode
  `\&12\catcode `\#12\catcode `\^12\catcode `\_12\catcode `\%12\relax}%
\providecommand \@@startlink[1]{}%
\providecommand \@@endlink[0]{}%
\providecommand \url  [0]{\begingroup\@sanitize@url \@url }%
\providecommand \@url [1]{\endgroup\@href {#1}{\urlprefix }}%
\providecommand \urlprefix  [0]{URL }%
\providecommand \Eprint [0]{\href }%
\providecommand \doibase [0]{https://doi.org/}%
\providecommand \selectlanguage [0]{\@gobble}%
\providecommand \bibinfo  [0]{\@secondoftwo}%
\providecommand \bibfield  [0]{\@secondoftwo}%
\providecommand \translation [1]{[#1]}%
\providecommand \BibitemOpen [0]{}%
\providecommand \bibitemStop [0]{}%
\providecommand \bibitemNoStop [0]{.\EOS\space}%
\providecommand \EOS [0]{\spacefactor3000\relax}%
\providecommand \BibitemShut  [1]{\csname bibitem#1\endcsname}%
\let\auto@bib@innerbib\@empty
%</preamble>
\bibitem [{\citenamefont {Lumley}(1967)}]{lumley1967structure}%
  \BibitemOpen
  \bibfield  {author} {\bibinfo {author} {\bibfnamefont {J.~L.}\ \bibnamefont
  {Lumley}},\ }\bibfield  {title} {\bibinfo {title} {The structure of
  inhomogeneous turbulent flows},\ }\href@noop {} {\bibfield  {journal}
  {\bibinfo  {journal} {Atmospheric turbulence and radio wave propagation}\ ,\
  \bibinfo {pages} {166}} (\bibinfo {year} {1967})}\BibitemShut {NoStop}%
\bibitem [{\citenamefont {Lumley}(1970)}]{lumley1970stochastic}%
  \BibitemOpen
  \bibfield  {author} {\bibinfo {author} {\bibfnamefont {J.~L.}\ \bibnamefont
  {Lumley}},\ }\href@noop {} {\emph {\bibinfo {title} {Stochastic tools in
  turbulence}}}\ (\bibinfo  {publisher} {Academic Press},\ \bibinfo {year}
  {1970})\BibitemShut {NoStop}%
\bibitem [{\citenamefont {Towne}\ \emph {et~al.}(2018)\citenamefont {Towne},
  \citenamefont {Schmidt},\ and\ \citenamefont {Colonius}}]{towne2018spectral}%
  \BibitemOpen
  \bibfield  {author} {\bibinfo {author} {\bibfnamefont {A.}~\bibnamefont
  {Towne}}, \bibinfo {author} {\bibfnamefont {O.~T.}\ \bibnamefont {Schmidt}},\
  and\ \bibinfo {author} {\bibfnamefont {T.}~\bibnamefont {Colonius}},\
  }\bibfield  {title} {\bibinfo {title} {Spectral proper orthogonal
  decomposition and its relationship to dynamic mode decomposition and
  resolvent analysis},\ }\href@noop {} {\bibfield  {journal} {\bibinfo
  {journal} {Journal of Fluid Mechanics}\ }\textbf {\bibinfo {volume} {847}},\
  \bibinfo {pages} {821} (\bibinfo {year} {2018})}\BibitemShut {NoStop}%
\bibitem [{\citenamefont {Rowley}\ \emph {et~al.}(2009)\citenamefont {Rowley},
  \citenamefont {Mezic}, \citenamefont {Bagheri}, \citenamefont {Schlatter},\
  and\ \citenamefont {Henningson}}]{Rowley2009jfm}%
  \BibitemOpen
  \bibfield  {author} {\bibinfo {author} {\bibfnamefont {C.~W.}\ \bibnamefont
  {Rowley}}, \bibinfo {author} {\bibfnamefont {I.}~\bibnamefont {Mezic}},
  \bibinfo {author} {\bibfnamefont {S.}~\bibnamefont {Bagheri}}, \bibinfo
  {author} {\bibfnamefont {P.}~\bibnamefont {Schlatter}},\ and\ \bibinfo
  {author} {\bibfnamefont {D.}~\bibnamefont {Henningson}},\ }\bibfield  {title}
  {\bibinfo {title} {Spectral analysis of nonlinear flows},\ }\href@noop {}
  {\bibfield  {journal} {\bibinfo  {journal} {J. Fluid Mech.}\ }\textbf
  {\bibinfo {volume} {645}},\ \bibinfo {pages} {115} (\bibinfo {year}
  {2009})}\BibitemShut {NoStop}%
\bibitem [{\citenamefont {Schmid}(2010)}]{schmid2010dynamic}%
  \BibitemOpen
  \bibfield  {author} {\bibinfo {author} {\bibfnamefont {P.~J.}\ \bibnamefont
  {Schmid}},\ }\bibfield  {title} {\bibinfo {title} {Dynamic mode decomposition
  of numerical and experimental data},\ }\href@noop {} {\bibfield  {journal}
  {\bibinfo  {journal} {Journal of fluid mechanics}\ }\textbf {\bibinfo
  {volume} {656}},\ \bibinfo {pages} {5} (\bibinfo {year} {2010})}\BibitemShut
  {NoStop}%
\bibitem [{\citenamefont {Schmid}(2022)}]{schmid2022dynamic}%
  \BibitemOpen
  \bibfield  {author} {\bibinfo {author} {\bibfnamefont {P.~J.}\ \bibnamefont
  {Schmid}},\ }\bibfield  {title} {\bibinfo {title} {Dynamic mode decomposition
  and its variants},\ }\href@noop {} {\bibfield  {journal} {\bibinfo  {journal}
  {Annual Review of Fluid Mechanics}\ }\textbf {\bibinfo {volume} {54}},\
  \bibinfo {pages} {225} (\bibinfo {year} {2022})}\BibitemShut {NoStop}%
\bibitem [{\citenamefont {Tu}\ \emph {et~al.}(2014{\natexlab{a}})\citenamefont
  {Tu}, \citenamefont {Rowley}, \citenamefont {Luchtenburg}, \citenamefont
  {Brunton},\ and\ \citenamefont {Kutz}}]{tu2014dmd}%
  \BibitemOpen
  \bibfield  {author} {\bibinfo {author} {\bibfnamefont {J.~H.}\ \bibnamefont
  {Tu}}, \bibinfo {author} {\bibfnamefont {C.~W.}\ \bibnamefont {Rowley}},
  \bibinfo {author} {\bibfnamefont {D.~M.}\ \bibnamefont {Luchtenburg}},
  \bibinfo {author} {\bibfnamefont {S.~L.}\ \bibnamefont {Brunton}},\ and\
  \bibinfo {author} {\bibfnamefont {J.~N.}\ \bibnamefont {Kutz}},\ }\bibfield
  {title} {\bibinfo {title} {On dynamic mode decomposition: Theory and
  applications},\ }\href@noop {} {\bibfield  {journal} {\bibinfo  {journal}
  {Journal of Computational Dynamics}\ }\textbf {\bibinfo {volume} {1}},\
  \bibinfo {pages} {391} (\bibinfo {year} {2014}{\natexlab{a}})}\BibitemShut
  {NoStop}%
\bibitem [{\citenamefont {Bridges}\ and\ \citenamefont
  {Wernet}(2003)}]{bridges2003measurements}%
  \BibitemOpen
  \bibfield  {author} {\bibinfo {author} {\bibfnamefont {J.}~\bibnamefont
  {Bridges}}\ and\ \bibinfo {author} {\bibfnamefont {M.}~\bibnamefont
  {Wernet}},\ }\bibfield  {title} {\bibinfo {title} {Measurements of
  aeroacoustic sound sources in turbulent jets},\ }in\ \href@noop {} {\emph
  {\bibinfo {booktitle} {9th AIAA/CEAS aeroacoustics conference and exhibit}}}\
  (\bibinfo {year} {2003})\ p.\ \bibinfo {pages} {3130}\BibitemShut {NoStop}%
\bibitem [{\citenamefont {Taira}\ \emph {et~al.}(2017)\citenamefont {Taira},
  \citenamefont {Brunton}, \citenamefont {Dawson}, \citenamefont {Rowley},
  \citenamefont {Colonius}, \citenamefont {McKeon}, \citenamefont {Schmidt},
  \citenamefont {Gordeyev}, \citenamefont {Theofilis},\ and\ \citenamefont
  {Ukeiley}}]{taira2017modal}%
  \BibitemOpen
  \bibfield  {author} {\bibinfo {author} {\bibfnamefont {K.}~\bibnamefont
  {Taira}}, \bibinfo {author} {\bibfnamefont {S.~L.}\ \bibnamefont {Brunton}},
  \bibinfo {author} {\bibfnamefont {S.~T.}\ \bibnamefont {Dawson}}, \bibinfo
  {author} {\bibfnamefont {C.~W.}\ \bibnamefont {Rowley}}, \bibinfo {author}
  {\bibfnamefont {T.}~\bibnamefont {Colonius}}, \bibinfo {author}
  {\bibfnamefont {B.~J.}\ \bibnamefont {McKeon}}, \bibinfo {author}
  {\bibfnamefont {O.~T.}\ \bibnamefont {Schmidt}}, \bibinfo {author}
  {\bibfnamefont {S.}~\bibnamefont {Gordeyev}}, \bibinfo {author}
  {\bibfnamefont {V.}~\bibnamefont {Theofilis}},\ and\ \bibinfo {author}
  {\bibfnamefont {L.~S.}\ \bibnamefont {Ukeiley}},\ }\bibfield  {title}
  {\bibinfo {title} {Modal analysis of fluid flows: An overview},\ }\href@noop
  {} {\bibfield  {journal} {\bibinfo  {journal} {Aiaa Journal}\ }\textbf
  {\bibinfo {volume} {55}},\ \bibinfo {pages} {4013} (\bibinfo {year}
  {2017})}\BibitemShut {NoStop}%
\bibitem [{\citenamefont {Tu}\ \emph {et~al.}(2014{\natexlab{b}})\citenamefont
  {Tu}, \citenamefont {Rowley}, \citenamefont {Kutz},\ and\ \citenamefont
  {Shang}}]{tu2014spectral}%
  \BibitemOpen
  \bibfield  {author} {\bibinfo {author} {\bibfnamefont {J.~H.}\ \bibnamefont
  {Tu}}, \bibinfo {author} {\bibfnamefont {C.~W.}\ \bibnamefont {Rowley}},
  \bibinfo {author} {\bibfnamefont {J.~N.}\ \bibnamefont {Kutz}},\ and\
  \bibinfo {author} {\bibfnamefont {J.~K.}\ \bibnamefont {Shang}},\ }\bibfield
  {title} {\bibinfo {title} {Spectral analysis of fluid flows using
  sub-nyquist-rate piv data},\ }\href@noop {} {\bibfield  {journal} {\bibinfo
  {journal} {Experiments in Fluids}\ }\textbf {\bibinfo {volume} {55}},\
  \bibinfo {pages} {1} (\bibinfo {year} {2014}{\natexlab{b}})}\BibitemShut
  {NoStop}%
\bibitem [{\citenamefont {Nekkanti}\ and\ \citenamefont
  {Schmidt}(2023)}]{nekkanti2023gappy}%
  \BibitemOpen
  \bibfield  {author} {\bibinfo {author} {\bibfnamefont {A.}~\bibnamefont
  {Nekkanti}}\ and\ \bibinfo {author} {\bibfnamefont {O.~T.}\ \bibnamefont
  {Schmidt}},\ }\bibfield  {title} {\bibinfo {title} {Gappy spectral proper
  orthogonal decomposition},\ }\href@noop {} {\bibfield  {journal} {\bibinfo
  {journal} {Journal of Computational Physics}\ }\textbf {\bibinfo {volume}
  {478}},\ \bibinfo {pages} {111950} (\bibinfo {year} {2023})}\BibitemShut
  {NoStop}%
\bibitem [{\citenamefont {Lee}\ \emph {et~al.}(2024)\citenamefont {Lee},
  \citenamefont {Ozawa}, \citenamefont {Nagata}, \citenamefont {Colonius},\
  and\ \citenamefont {Nonomura}}]{lee2024superresolution}%
  \BibitemOpen
  \bibfield  {author} {\bibinfo {author} {\bibfnamefont {C.}~\bibnamefont
  {Lee}}, \bibinfo {author} {\bibfnamefont {Y.}~\bibnamefont {Ozawa}}, \bibinfo
  {author} {\bibfnamefont {T.}~\bibnamefont {Nagata}}, \bibinfo {author}
  {\bibfnamefont {T.}~\bibnamefont {Colonius}},\ and\ \bibinfo {author}
  {\bibfnamefont {T.}~\bibnamefont {Nonomura}},\ }\bibfield  {title} {\bibinfo
  {title} {Superresolution and analysis of three-dimensional velocity fields of
  underexpanded jets in different screech modes},\ }\href@noop {} {\bibfield
  {journal} {\bibinfo  {journal} {Physical Review Fluids}\ }\textbf {\bibinfo
  {volume} {9}},\ \bibinfo {pages} {104604} (\bibinfo {year}
  {2024})}\BibitemShut {NoStop}%
\bibitem [{\citenamefont {Zhang}\ \emph {et~al.}(2020)\citenamefont {Zhang},
  \citenamefont {Cattafesta},\ and\ \citenamefont
  {Ukeiley}}]{zhang2020spectral}%
  \BibitemOpen
  \bibfield  {author} {\bibinfo {author} {\bibfnamefont {Y.}~\bibnamefont
  {Zhang}}, \bibinfo {author} {\bibfnamefont {L.~N.}\ \bibnamefont
  {Cattafesta}},\ and\ \bibinfo {author} {\bibfnamefont {L.}~\bibnamefont
  {Ukeiley}},\ }\bibfield  {title} {\bibinfo {title} {Spectral analysis modal
  methods (samms) using non-time-resolved piv},\ }\href@noop {} {\bibfield
  {journal} {\bibinfo  {journal} {Experiments in Fluids}\ }\textbf {\bibinfo
  {volume} {61}},\ \bibinfo {pages} {1} (\bibinfo {year} {2020})}\BibitemShut
  {NoStop}%
\bibitem [{\citenamefont {Schmidt}\ and\ \citenamefont
  {Colonius}(2020)}]{schmidt2020guide}%
  \BibitemOpen
  \bibfield  {author} {\bibinfo {author} {\bibfnamefont {O.~T.}\ \bibnamefont
  {Schmidt}}\ and\ \bibinfo {author} {\bibfnamefont {T.}~\bibnamefont
  {Colonius}},\ }\bibfield  {title} {\bibinfo {title} {Guide to spectral proper
  orthogonal decomposition},\ }\href@noop {} {\bibfield  {journal} {\bibinfo
  {journal} {Aiaa journal}\ }\textbf {\bibinfo {volume} {58}},\ \bibinfo
  {pages} {1023} (\bibinfo {year} {2020})}\BibitemShut {NoStop}%
\bibitem [{\citenamefont {Heidt}\ and\ \citenamefont
  {Colonius}(2024)}]{heidt2024optimal}%
  \BibitemOpen
  \bibfield  {author} {\bibinfo {author} {\bibfnamefont {L.}~\bibnamefont
  {Heidt}}\ and\ \bibinfo {author} {\bibfnamefont {T.}~\bibnamefont
  {Colonius}},\ }\bibfield  {title} {\bibinfo {title} {Optimal frequency
  resolution for spectral proper orthogonal decomposition},\ }\href@noop {}
  {\bibfield  {journal} {\bibinfo  {journal} {arXiv preprint arXiv:2402.15775}\
  } (\bibinfo {year} {2024})}\BibitemShut {NoStop}%
\bibitem [{\citenamefont {Chen}\ \emph {et~al.}(2012)\citenamefont {Chen},
  \citenamefont {Tu},\ and\ \citenamefont {Rowley}}]{chen2012variants}%
  \BibitemOpen
  \bibfield  {author} {\bibinfo {author} {\bibfnamefont {K.~K.}\ \bibnamefont
  {Chen}}, \bibinfo {author} {\bibfnamefont {J.~H.}\ \bibnamefont {Tu}},\ and\
  \bibinfo {author} {\bibfnamefont {C.~W.}\ \bibnamefont {Rowley}},\ }\bibfield
   {title} {\bibinfo {title} {Variants of dynamic mode decomposition: boundary
  condition, koopman, and fourier analyses},\ }\href@noop {} {\bibfield
  {journal} {\bibinfo  {journal} {Journal of nonlinear science}\ }\textbf
  {\bibinfo {volume} {22}},\ \bibinfo {pages} {887} (\bibinfo {year}
  {2012})}\BibitemShut {NoStop}%
\bibitem [{\citenamefont {Bagheri}(2014)}]{bagheri2014effects}%
  \BibitemOpen
  \bibfield  {author} {\bibinfo {author} {\bibfnamefont {S.}~\bibnamefont
  {Bagheri}},\ }\bibfield  {title} {\bibinfo {title} {Effects of weak noise on
  oscillating flows: Linking quality factor, floquet modes, and koopman
  spectrum},\ }\href@noop {} {\bibfield  {journal} {\bibinfo  {journal}
  {Physics of Fluids}\ }\textbf {\bibinfo {volume} {26}} (\bibinfo {year}
  {2014})}\BibitemShut {NoStop}%
\bibitem [{\citenamefont {Jovanovi{\'c}}\ \emph {et~al.}(2014)\citenamefont
  {Jovanovi{\'c}}, \citenamefont {Schmid},\ and\ \citenamefont
  {Nichols}}]{Jovanovic2014pof}%
  \BibitemOpen
  \bibfield  {author} {\bibinfo {author} {\bibfnamefont {M.~R.}\ \bibnamefont
  {Jovanovi{\'c}}}, \bibinfo {author} {\bibfnamefont {P.~J.}\ \bibnamefont
  {Schmid}},\ and\ \bibinfo {author} {\bibfnamefont {J.~W.}\ \bibnamefont
  {Nichols}},\ }\bibfield  {title} {\bibinfo {title} {Sparsity-promoting
  dynamic mode decomposition},\ }\href@noop {} {\bibfield  {journal} {\bibinfo
  {journal} {Physics of Fluids}\ }\textbf {\bibinfo {volume} {26}},\ \bibinfo
  {pages} {024103} (\bibinfo {year} {2014})}\BibitemShut {NoStop}%
\bibitem [{\citenamefont {Hemati}\ \emph {et~al.}(2017)\citenamefont {Hemati},
  \citenamefont {Rowley}, \citenamefont {Deem},\ and\ \citenamefont
  {Cattafesta}}]{hemati2017biasing}%
  \BibitemOpen
  \bibfield  {author} {\bibinfo {author} {\bibfnamefont {M.~S.}\ \bibnamefont
  {Hemati}}, \bibinfo {author} {\bibfnamefont {C.~W.}\ \bibnamefont {Rowley}},
  \bibinfo {author} {\bibfnamefont {E.~A.}\ \bibnamefont {Deem}},\ and\
  \bibinfo {author} {\bibfnamefont {L.~N.}\ \bibnamefont {Cattafesta}},\
  }\bibfield  {title} {\bibinfo {title} {De-biasing the dynamic mode
  decomposition for applied koopman spectral analysis of noisy datasets},\
  }\href@noop {} {\bibfield  {journal} {\bibinfo  {journal} {Theoretical and
  Computational Fluid Dynamics}\ }\textbf {\bibinfo {volume} {31}},\ \bibinfo
  {pages} {349} (\bibinfo {year} {2017})}\BibitemShut {NoStop}%
\bibitem [{\citenamefont {Askham}\ and\ \citenamefont
  {Kutz}(2018)}]{askham2018variable}%
  \BibitemOpen
  \bibfield  {author} {\bibinfo {author} {\bibfnamefont {T.}~\bibnamefont
  {Askham}}\ and\ \bibinfo {author} {\bibfnamefont {J.~N.}\ \bibnamefont
  {Kutz}},\ }\bibfield  {title} {\bibinfo {title} {Variable projection methods
  for an optimized dynamic mode decomposition},\ }\href@noop {} {\bibfield
  {journal} {\bibinfo  {journal} {SIAM Journal on Applied Dynamical Systems}\
  }\textbf {\bibinfo {volume} {17}},\ \bibinfo {pages} {380} (\bibinfo {year}
  {2018})}\BibitemShut {NoStop}%
\bibitem [{\citenamefont {Baddoo}\ \emph {et~al.}(2023)\citenamefont {Baddoo},
  \citenamefont {Herrmann}, \citenamefont {McKeon}, \citenamefont
  {Nathan~Kutz},\ and\ \citenamefont {Brunton}}]{baddoo2023physics}%
  \BibitemOpen
  \bibfield  {author} {\bibinfo {author} {\bibfnamefont {P.~J.}\ \bibnamefont
  {Baddoo}}, \bibinfo {author} {\bibfnamefont {B.}~\bibnamefont {Herrmann}},
  \bibinfo {author} {\bibfnamefont {B.~J.}\ \bibnamefont {McKeon}}, \bibinfo
  {author} {\bibfnamefont {J.}~\bibnamefont {Nathan~Kutz}},\ and\ \bibinfo
  {author} {\bibfnamefont {S.~L.}\ \bibnamefont {Brunton}},\ }\bibfield
  {title} {\bibinfo {title} {Physics-informed dynamic mode decomposition},\
  }\href@noop {} {\bibfield  {journal} {\bibinfo  {journal} {Proceedings of the
  Royal Society A}\ }\textbf {\bibinfo {volume} {479}},\ \bibinfo {pages}
  {20220576} (\bibinfo {year} {2023})}\BibitemShut {NoStop}%
\bibitem [{\citenamefont {Dawson}\ \emph {et~al.}(2016)\citenamefont {Dawson},
  \citenamefont {Hemati}, \citenamefont {Williams},\ and\ \citenamefont
  {Rowley}}]{dawson2016characterizing}%
  \BibitemOpen
  \bibfield  {author} {\bibinfo {author} {\bibfnamefont {S.~T.}\ \bibnamefont
  {Dawson}}, \bibinfo {author} {\bibfnamefont {M.~S.}\ \bibnamefont {Hemati}},
  \bibinfo {author} {\bibfnamefont {M.~O.}\ \bibnamefont {Williams}},\ and\
  \bibinfo {author} {\bibfnamefont {C.~W.}\ \bibnamefont {Rowley}},\ }\bibfield
   {title} {\bibinfo {title} {Characterizing and correcting for the effect of
  sensor noise in the dynamic mode decomposition},\ }\href@noop {} {\bibfield
  {journal} {\bibinfo  {journal} {Experiments in Fluids}\ }\textbf {\bibinfo
  {volume} {57}},\ \bibinfo {pages} {1} (\bibinfo {year} {2016})}\BibitemShut
  {NoStop}%
\bibitem [{\citenamefont {Li}\ \emph {et~al.}(2022)\citenamefont {Li},
  \citenamefont {Garicano-Mena},\ and\ \citenamefont {Valero}}]{li2022dynamic}%
  \BibitemOpen
  \bibfield  {author} {\bibinfo {author} {\bibfnamefont {B.}~\bibnamefont
  {Li}}, \bibinfo {author} {\bibfnamefont {J.}~\bibnamefont {Garicano-Mena}},\
  and\ \bibinfo {author} {\bibfnamefont {E.}~\bibnamefont {Valero}},\
  }\bibfield  {title} {\bibinfo {title} {A dynamic mode decomposition technique
  for the analysis of non--uniformly sampled flow data},\ }\href@noop {}
  {\bibfield  {journal} {\bibinfo  {journal} {Journal of Computational
  Physics}\ }\textbf {\bibinfo {volume} {468}},\ \bibinfo {pages} {111495}
  (\bibinfo {year} {2022})}\BibitemShut {NoStop}%
\bibitem [{\citenamefont {Kaneko}\ \emph {et~al.}(2024)\citenamefont {Kaneko},
  \citenamefont {del Pozo}, \citenamefont {Nishikori}, \citenamefont {Ozawa},\
  and\ \citenamefont {Nonomura}}]{kaneko2024dmd}%
  \BibitemOpen
  \bibfield  {author} {\bibinfo {author} {\bibfnamefont {S.}~\bibnamefont
  {Kaneko}}, \bibinfo {author} {\bibfnamefont {A.}~\bibnamefont {del Pozo}},
  \bibinfo {author} {\bibfnamefont {H.}~\bibnamefont {Nishikori}}, \bibinfo
  {author} {\bibfnamefont {Y.}~\bibnamefont {Ozawa}},\ and\ \bibinfo {author}
  {\bibfnamefont {T.}~\bibnamefont {Nonomura}},\ }\bibfield  {title} {\bibinfo
  {title} {Dmd-based spatiotemporal superresolution measurement of a supersonic
  jet using dual planar piv and acoustic data},\ }\href@noop {} {\bibfield
  {journal} {\bibinfo  {journal} {Experiments in Fluids}\ }\textbf {\bibinfo
  {volume} {65}},\ \bibinfo {pages} {139} (\bibinfo {year} {2024})}\BibitemShut
  {NoStop}%
\bibitem [{\citenamefont {Bagheri}\ \emph {et~al.}(2009)\citenamefont
  {Bagheri}, \citenamefont {Henningson}, \citenamefont {Hoepffner},\ and\
  \citenamefont {Schmid}}]{bagheri2009input}%
  \BibitemOpen
  \bibfield  {author} {\bibinfo {author} {\bibfnamefont {S.}~\bibnamefont
  {Bagheri}}, \bibinfo {author} {\bibfnamefont {D.~S.}\ \bibnamefont
  {Henningson}}, \bibinfo {author} {\bibfnamefont {J.}~\bibnamefont
  {Hoepffner}},\ and\ \bibinfo {author} {\bibfnamefont {P.~J.}\ \bibnamefont
  {Schmid}},\ }\bibfield  {title} {\bibinfo {title} {Input-output analysis and
  control design applied to a linear model of spatially developing flows},\
  }\href@noop {} {\bibfield  {journal} {\bibinfo  {journal} {Applied Mechanics
  Reviews}\ }\textbf {\bibinfo {volume} {62}} (\bibinfo {year}
  {2009})}\BibitemShut {NoStop}%
\bibitem [{\citenamefont {Hirsh}\ \emph {et~al.}(2020)\citenamefont {Hirsh},
  \citenamefont {Harris}, \citenamefont {Kutz},\ and\ \citenamefont
  {Brunton}}]{hirsh2020centering}%
  \BibitemOpen
  \bibfield  {author} {\bibinfo {author} {\bibfnamefont {S.~M.}\ \bibnamefont
  {Hirsh}}, \bibinfo {author} {\bibfnamefont {K.~D.}\ \bibnamefont {Harris}},
  \bibinfo {author} {\bibfnamefont {J.~N.}\ \bibnamefont {Kutz}},\ and\
  \bibinfo {author} {\bibfnamefont {B.~W.}\ \bibnamefont {Brunton}},\
  }\bibfield  {title} {\bibinfo {title} {Centering data improves the dynamic
  mode decomposition},\ }\href@noop {} {\bibfield  {journal} {\bibinfo
  {journal} {SIAM Journal on Applied Dynamical Systems}\ }\textbf {\bibinfo
  {volume} {19}},\ \bibinfo {pages} {1920} (\bibinfo {year}
  {2020})}\BibitemShut {NoStop}%
\bibitem [{\citenamefont {Scherl}\ \emph {et~al.}(2020)\citenamefont {Scherl},
  \citenamefont {Strom}, \citenamefont {Shang}, \citenamefont {Williams},
  \citenamefont {Polagye},\ and\ \citenamefont {Brunton}}]{scherl2020robust}%
  \BibitemOpen
  \bibfield  {author} {\bibinfo {author} {\bibfnamefont {I.}~\bibnamefont
  {Scherl}}, \bibinfo {author} {\bibfnamefont {B.}~\bibnamefont {Strom}},
  \bibinfo {author} {\bibfnamefont {J.~K.}\ \bibnamefont {Shang}}, \bibinfo
  {author} {\bibfnamefont {O.}~\bibnamefont {Williams}}, \bibinfo {author}
  {\bibfnamefont {B.~L.}\ \bibnamefont {Polagye}},\ and\ \bibinfo {author}
  {\bibfnamefont {S.~L.}\ \bibnamefont {Brunton}},\ }\bibfield  {title}
  {\bibinfo {title} {Robust principal component analysis for modal
  decomposition of corrupt fluid flows},\ }\href@noop {} {\bibfield  {journal}
  {\bibinfo  {journal} {Physical Review Fluids}\ }\textbf {\bibinfo {volume}
  {5}},\ \bibinfo {pages} {054401} (\bibinfo {year} {2020})}\BibitemShut
  {NoStop}%
\bibitem [{\citenamefont {Br{\`e}s}\ \emph {et~al.}(2017)\citenamefont
  {Br{\`e}s}, \citenamefont {Ham}, \citenamefont {Nichols},\ and\ \citenamefont
  {Lele}}]{bres2017unstructured}%
  \BibitemOpen
  \bibfield  {author} {\bibinfo {author} {\bibfnamefont {G.~A.}\ \bibnamefont
  {Br{\`e}s}}, \bibinfo {author} {\bibfnamefont {F.~E.}\ \bibnamefont {Ham}},
  \bibinfo {author} {\bibfnamefont {J.~W.}\ \bibnamefont {Nichols}},\ and\
  \bibinfo {author} {\bibfnamefont {S.~K.}\ \bibnamefont {Lele}},\ }\bibfield
  {title} {\bibinfo {title} {Unstructured large-eddy simulations of supersonic
  jets},\ }\href@noop {} {\bibfield  {journal} {\bibinfo  {journal} {AIAA
  journal}\ }\textbf {\bibinfo {volume} {55}},\ \bibinfo {pages} {1164}
  (\bibinfo {year} {2017})}\BibitemShut {NoStop}%
\bibitem [{\citenamefont {Br{\`e}s}\ \emph {et~al.}(2018)\citenamefont
  {Br{\`e}s}, \citenamefont {Jordan}, \citenamefont {Jaunet}, \citenamefont
  {Le~Rallic}, \citenamefont {Cavalieri}, \citenamefont {Towne}, \citenamefont
  {Lele}, \citenamefont {Colonius},\ and\ \citenamefont
  {Schmidt}}]{bres2018importance}%
  \BibitemOpen
  \bibfield  {author} {\bibinfo {author} {\bibfnamefont {G.~A.}\ \bibnamefont
  {Br{\`e}s}}, \bibinfo {author} {\bibfnamefont {P.}~\bibnamefont {Jordan}},
  \bibinfo {author} {\bibfnamefont {V.}~\bibnamefont {Jaunet}}, \bibinfo
  {author} {\bibfnamefont {M.}~\bibnamefont {Le~Rallic}}, \bibinfo {author}
  {\bibfnamefont {A.~V.}\ \bibnamefont {Cavalieri}}, \bibinfo {author}
  {\bibfnamefont {A.}~\bibnamefont {Towne}}, \bibinfo {author} {\bibfnamefont
  {S.~K.}\ \bibnamefont {Lele}}, \bibinfo {author} {\bibfnamefont
  {T.}~\bibnamefont {Colonius}},\ and\ \bibinfo {author} {\bibfnamefont
  {O.~T.}\ \bibnamefont {Schmidt}},\ }\bibfield  {title} {\bibinfo {title}
  {Importance of the nozzle-exit boundary-layer state in subsonic turbulent
  jets},\ }\href@noop {} {\bibfield  {journal} {\bibinfo  {journal} {Journal of
  Fluid Mechanics}\ }\textbf {\bibinfo {volume} {851}},\ \bibinfo {pages} {83}
  (\bibinfo {year} {2018})}\BibitemShut {NoStop}%
\bibitem [{\citenamefont {Chu}(1965)}]{chu1965energy}%
  \BibitemOpen
  \bibfield  {author} {\bibinfo {author} {\bibfnamefont {B.-T.}\ \bibnamefont
  {Chu}},\ }\bibfield  {title} {\bibinfo {title} {On the energy transfer to
  small disturbances in fluid flow (part i)},\ }\href@noop {} {\bibfield
  {journal} {\bibinfo  {journal} {Acta Mechanica}\ }\textbf {\bibinfo {volume}
  {1}},\ \bibinfo {pages} {215} (\bibinfo {year} {1965})}\BibitemShut {NoStop}%
\bibitem [{\citenamefont {Asztalos}\ \emph {et~al.}(2023)\citenamefont
  {Asztalos}, \citenamefont {Almashjary},\ and\ \citenamefont
  {Dawson}}]{asztalos2023galerkin}%
  \BibitemOpen
  \bibfield  {author} {\bibinfo {author} {\bibfnamefont {K.~J.}\ \bibnamefont
  {Asztalos}}, \bibinfo {author} {\bibfnamefont {A.}~\bibnamefont
  {Almashjary}},\ and\ \bibinfo {author} {\bibfnamefont {S.~T.}\ \bibnamefont
  {Dawson}},\ }\bibfield  {title} {\bibinfo {title} {Galerkin spectral
  estimation of vortex-dominated wake flows},\ }\href@noop {} {\bibfield
  {journal} {\bibinfo  {journal} {Theoretical and Computational Fluid
  Dynamics}\ ,\ \bibinfo {pages} {1}} (\bibinfo {year} {2023})}\BibitemShut
  {NoStop}%
\bibitem [{\citenamefont {Schmidt}(2020)}]{schmidt2020bispectral}%
  \BibitemOpen
  \bibfield  {author} {\bibinfo {author} {\bibfnamefont {O.~T.}\ \bibnamefont
  {Schmidt}},\ }\bibfield  {title} {\bibinfo {title} {Bispectral mode
  decomposition of nonlinear flows},\ }\href@noop {} {\bibfield  {journal}
  {\bibinfo  {journal} {Nonlinear Dynamics}\ }\textbf {\bibinfo {volume}
  {102}},\ \bibinfo {pages} {2479} (\bibinfo {year} {2020})}\BibitemShut
  {NoStop}%
\end{thebibliography}%


%apsrev4-2.bst 2019-01-14 (MD) hand-edited version of apsrev4-1.bst
%Control: key (0)
%Control: author (8) initials jnrlst
%Control: editor formatted (1) identically to author
%Control: production of article title (0) allowed
%Control: page (0) single
%Control: year (1) truncated
%Control: production of eprint (0) enabled
%

\end{document}